\documentclass[twocolumn]{jpsj3}
\usepackage{txfonts}

\title{Magnetic, Transport, and Phonon Properties of the Trivalent Eu Metallic Compound EuBe$_{13}$}

\author{Hiroyuki Hidaka\thanks{E-mail: hidaka@phys.sci.hokudai.ac.jp}, Kota Mizuuchi, Tatsuya Yanagisawa, and Hiroshi Amitsuka}

\inst{Graduate School of Science, Hokkaido University, Sapporo, Hokkaido 060-0810, Japan} %\\

\abst{
Magnetic susceptibility $\chi$, specific heat $C$, and electrical resistivity $\rho$ measurements have been performed on single-crystal EuBe$_{13}$ in the temperature range between 2 and 300 K to investigate its phonon property and the valence state of the Eu ion. 
The obtained $\chi$($T$) curves obey a typical Van Vleck susceptibility for Eu$^{3+}$ with a nonmagnetic ground state in the entire measured temperature range. 
In the case of $C$($T$), we observed the coexistence of Debye and Einstein phonon modes with characteristic Debye and Einstein temperatures of $\theta_{\rm D}$ $\sim$ 835 K and $\theta_{\rm E}$ $\sim$ 167 K, respectively, which are in good agreement with those previously reported for other isostructural MBe$_{13}$ compounds (M = rare earths and actinides). 
The $\rho$($T$) curve for EuBe$_{13}$ shows an unusual $T^3$-like dependence at low temperatures, as also observed for the nonmagnetic isostructural compound LaBe$_{13}$, which can be reproduced well by calculations based on electron--phonon scattering using the estimated $\theta_{\rm D}$ and $\theta_{\rm E}$. 
We also summarized the relationship between the Eu-valence state and the free distance between the Eu ion and the first-nearest-neighbor atoms in several Eu-based cubic compounds, and argued that EuBe$_{13}$ takes the Eu$^{3+}$ state despite its larger free distance than other Eu$^{3+}$ compounds. } 

\begin{document}
\maketitle

\section{Introduction} 
MBe$_{13}$ compounds (M = rare earths and actinides) show a rich variety of physical properties depending on the M ion, such as unconventional superconductivity (SC) and non-Fermi-liquid behavior in UBe$_{13}$ \cite{Ott, Mayer}, an intermediate valence state in CeBe$_{13}$ \cite{Wilson}, helical magnetic ordering in HoBe$_{13}$ \cite{Bouree}, and nuclear antiferromagnetic (AFM) ordering in PrBe$_{13}$ \cite{Moyland}. 
They crystallize in a NaZn$_{13}$-type cubic structure with the space group $F$$m$$\bar{\rm 3}$$c$ (No. 226, $O_h^{\rm 6}$), where the unit cell contains M atoms in the 8$a$ site, Be$^{\rm I}$ atoms in the 8$b$ site, and Be$^{\rm II}$ atoms in the 96$i$ site \cite{Bucher, McElfresh, Takegahara}. 
It is notable that the MBe$_{13}$ compounds can be categorized as cage-structured compounds, since the unit cell consists of two cagelike structures; the M atom is surrounded by 24 Be$^{\rm II}$ atoms, nearly forming a snub cube, and the Be$^{\rm I}$ atom is surrounded by 12 Be$^{\rm II}$ atoms, forming an icosahedron cage. 
Such cage-structured compounds have attracted much attention because of the presence of a low-energy phonon mode associated with local vibration of a guest atom with a large amplitude in an oversized host cage, so-called rattling \cite{Caplin, Mandrus, Matsuhira, Yamaura2, Suekuni, Yamaura1}. 
The low-energy phonon mode has been considered to be related to several intriguing phenomena, such as rattling-induced superconductivity \cite{Nagao} and a magnetic-field-insensitive heavy-fermion state \cite{Sanada, Hattori}, via electron--phonon coupling.

In several MBe$_{13}$ compounds, such as LaBe$_{13}$, SmBe$_{13}$, UBe$_{13}$, and ThBe$_{13}$, a low-energy phonon mode, which can be described well by a model assuming a conventional harmonic Einstein phonon, has also been observed \cite{Renker, Hidaka-La, Hidaka-XRD}. 
These findings suggest that the low-energy phonon mode is common to the MBe$_{13}$ compounds. 
Previous results of inelastic neutron scattering (INS) and powder X-ray diffraction (XRD) measurements strongly indicate that the M atom behaves as an Einstein oscillator with characteristic temperature $\theta_{\rm E}$ $\sim$ 160 K, whereas the Be atoms form the crystal lattice described by the Debye model with characteristic temperatures $\theta_{\rm D}$ $\sim$ 600 -- 800 K \cite{Renker, Hidaka-XRD, Kappler}. 
Interestingly, the obtained $\theta_{\rm E}$ values in these systems appear to be independent of either the mass of the guest atoms or the guest free distances in the snub cube, which is a characteristic feature not found in other cage-structured compounds having a similar low-energy phonon mode \cite{Suekuni, Matsuhira, Yamaura1, Yamaura2}. 
To obtain further insight into the characteristics of the low-energy phonon modes and their effects on the electronic states in MBe$_{13}$, it is necessary to explore the phonon and electronic properties in other isostructural MBe$_{13}$ compounds.

On the other hand, Eu-based compounds show two types of Eu valency: divalent (Eu$^{2+}$) and trivalent (Eu$^{3+}$). 
The ground state for Eu$^{2+}$ is magnetic (4$f$$^7$: $S$ = 7/2, $L$ = 0, and $J$ = 7/2), while Eu$^{3+}$ is nonmagnetic (4$f$$^6$: $S$ = 3, $L$ = 3, and $J$ = 0). 
Here, $S$, $L$, and $J$ are the total spin, total orbital, and total angular momenta, respectively. 
It is interesting that the number of intermetallic compounds with Eu$^{3+}$ found thus far at ambient pressure is much smaller than that with Eu$^{2+}$ or the intermediate valence state \cite{Onuki-Philo}, even though most rare-earth ions are usually trivalent in their compounds. 
The valence state of the Eu ion can be tuned easily by external parameters, such as temperature and pressure, because the energy difference between the two valence states is relatively small \cite{Hotta}.
In this context, it will be useful to examine the relationship between the valence state and free space of the Eu ion in various Eu-based compounds, since the effective radii of Eu ions are different between the two valence states. 

EuBe$_{13}$ is a valuable material for studying not only the phonon property in MBe$_{13}$ systems but also characteristics of the Eu$^{3+}$ state. 
Its Eu valence has been revealed to be trivalent from previous magnetic susceptibility ($\chi$) and M$\rm \ddot{o}$ssbauer spectroscopy measurements \cite{Bucher, Nowik}. 
However, these measurements were performed for polycrystalline samples, and no information about the low-energy phonon mode and fundamental physical properties except for $\chi$ was given \cite{Bucher, Nowik, ESR}. 
In this paper, we report the results of $\chi$, specific heat ($C$), and electrical resistivity ($\rho$) measurements on single-crystal EuBe$_{13}$, and provide evidence of the presence of the low-energy phonon mode and the pure trivalent state of the Eu ion.

\section{Experimental Procedure}
Single crystals of EuBe$_{13}$ were grown by the Al-flux method. 
The constituent materials (Eu with 99.9$\%$ purity and Be with 99.9$\%$ purity) and Al with 99.99$\%$ purity were placed in an Al$_2$O$_3$ crucible at an atomic ratio of 1:13:35 and sealed in a quartz tube filled with Ar gas of $\sim$ 150 mmHg. 
The sealed tube was kept at 1050 $\degC$ for 3 days and then cooled at a rate of 2 $\degC$/h. 
The Al flux was spun in a centrifuge and then removed using NaOH solution. 
The typical size of a grown sample is about 1 $\times$ 1 $\times$ 1 mm$^3$. 
The results of powder XRD measurement at room temperature showed no impurity phase within the experimental accuracy except for reflections from a copper holder, although $\chi$ measurements indicate that the present single crystals include a minute amount of magnetic impurities, which may come from some Eu--Al binary alloy. 
A lattice parameter of EuBe$_{13}$ was obtained to be $a$ = 10.299(1) $\AA$, which is close to the previously reported value of $a$ = 10.286 $\AA$ \cite{Bucher}.

The DC magnetization ($M$) was measured in the temperature range from 2 to 300 K at magnetic fields $B$ = 0.1 and 1 T using a Magnetic Property Measurement System (MPMS, Quantum Design, Inc.) and two crystal pieces (samples \#1 and \#2) taken from the same batch. 
$C$ was measured in the temperature range of 2 -- 300 K at 0 T with a Physical Property Measurement System (PPMS, Quantum Design, Inc.) using a crystal piece (sample \#3) taken from the same batch. 
$\rho$ was measured using sample \#1 by a conventional four-probe method in the temperature range of 1.3 -- 300 K at 0 T with a $^4$He refrigerator. 
The electrical current $\boldmath I$ was applied along the [100] direction.

\section{Experimental Results}

\subsection{Magnetic susceptibility}

Figure 1 shows the temperature dependence of the magnetic susceptibility $\chi$($T$) (= $M$($T$)/$B$) for sample \#1 of EuBe$_{13}$ measured at $B$ = 0.1 and 1 T between 2 and 300 K. 
The magnetic field was applied along the [100] axis. 
Both $\chi$($T$) curves gradually increase with decreasing temperature, and then become nearly constant below $\sim$ 100 K. 
Below $\sim$ 50 K, the $\chi$($T$) curves start to increase again and show a clear cusp at $\sim$ 15 K for $B$ = 0.1 T, while a broad shoulder appears for $B$ = 1 T. 
The increase in $\chi$($T$) at the low temperatures can also be observed in $\chi$($T$) at 0.1 T for sample \#2, as shown in the inset of Fig. 1, which is more prominent than that for sample \#1. 
These magnetic field and sample dependences indicate that the increase in $\chi$($T$) at low temperatures can be attributed to some magnetic impurities undetected in the XRD measurements. 
One of the possible impurities is the antiferromagnet EuAl$_4$ with Eu$^{2+}$ ($T_{\rm N}$= 15.4 K) \cite{Nakamura-EuAl4}, since the present $\chi$($T$) curves at 0.1 T show a cusp anomaly at $\sim$ 15 K. 
In addition, a further upturn below 10 K in $\chi$($T$), which cannot be explained by EuAl$_4$, indicates the presence of other magnetic impurities. 
We roughly estimated the amount of impurities in sample \#2 to be about 0.5$\%$, on the assumption that the low-$T$ increase in $\chi$($T$) at 0.1 T comes from EuAl$_4$ and some other Eu$^{2+}$ paramagnetic material.

The obtained $\chi$($T$) for EuBe$_{13}$ well obeys the Van Vleck paramagnetic susceptibility, except for the low-$T$ increase due to the minute amount of magnetic impurities. 
In the case of Eu$^{3+}$, the ground-state $J$ multiplet is $J$ = 0, and the energy of the excited $J$ multiplet $E_J$ is given as 
\begin{equation} 
E_J = \frac{\lambda}{2}J(J+1), 
\end{equation}
where $\lambda$ is the coupling constant of the spin-orbit interaction $\lambda$\mbox{\boldmath $L$}$\cdot$\mbox{\boldmath $S$}. 
The Van Vleck magnetic susceptibility $\chi_{\rm vv}$ can be expressed as \cite{VanVleck}: 
\begin{eqnarray} 
\chi_{\rm vv}(T) = \frac{\sum^{J=6}_{J=0}\chi_J(2J+1) e^{-E_J/k_{\rm B}T}} {\sum^{J=6}_{J=0} (2J+1) e^{-E_J/k_{\rm B}T} }, 
\end{eqnarray} 
where $\chi_J$ is written as 
\begin{eqnarray} 
\chi_J = \frac{N_{\rm A} g_J^2 \mu_{\rm B}^2 J(J+1)}{3k_{\rm B}T} + \alpha_J,
\end{eqnarray}
\begin{eqnarray} 
\alpha_J = \frac{N_{\rm A} \mu_{\rm B}^2 }{6(2J+1)} \biggl( \frac{F_{J+1}}{E_{J+1}-E_J} - \frac{F_J}{E_J-E_{J-1}} \biggl),
\end{eqnarray}
\begin{eqnarray} 
F_J = \frac{[(L+S+1)^2-J^2][J^2-(S-L)^2]}{J}. 
\end{eqnarray}
Here, $k_{\rm B}$ is the Boltzmann constant, $N_{\rm A}$ Avogadro's number, $\mu_{\rm B}$ the Bohr magneton, and $g_J$ the Land\'e g factor. 
The best fit of Eq. (2) to the experimental data, which is represented by the red solid line in Fig. 1, gives the value of $\lambda$ = 481 K. 
This is in good agreement with that obtained from the theoretical calculation on the assumption of free Eu$^{3+}$ ($\lambda$ = 460 K) \cite{Judd} and the previous experiment using a polycrystalline sample ($\lambda$ $\sim$ 476 K) \cite{Bucher}, indicating the pure Eu$^{3+}$ state of EuBe$_{13}$ at temperatures below 300 K. 

\begin{figure}[htb]
\begin{center}
\includegraphics[width=0.8\linewidth]{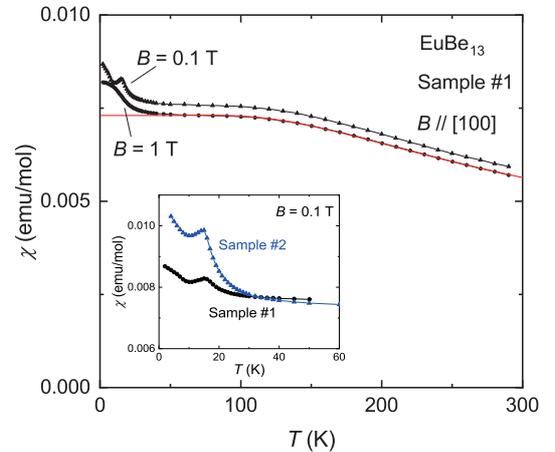}
\caption[]{\protect (Color online) Temperature dependence of the magnetic susceptibility $\chi$($T$) for sample \#1 of EuBe$_{13}$ at $B$ = 0.1 and 1 T ($B$ // [100]). 
The red solid line represents the fitting curve on the basis of the Van Vleck susceptibility, as described in the text. 
The inset shows $\chi$($T$) below 60 K for samples \#1 and \#2 at $B$ = 0.1 T. 
}
\end{center}
\end{figure}

\subsection{Specific heat}

Figure 2 shows the temperature dependence of the specific heat divided by the temperature $C$($T$)/$T$ for sample \#3 of EuBe$_{13}$. 
$C$($T$)/$T$ for LaBe$_{13}$ is also displayed in this figure for comparison \cite{Hidaka-La}. 
The $C$($T$)/$T$ curve for EuBe$_{13}$ is similar to that for LaBe$_{13}$, indicating that the contribution of 4$f$ electrons of the Eu ion to the specific heat is negligibly small. 
In addition, there is no indication of a phase transition near 15 K, where the cusp anomaly was observed in the $\chi(T)$ curve. 
As shown in the inset of Fig. 2, the $C$($T$)/$T$ curve obeys the Debye $T^3$ law below $\sim$ 13 K: $C$($T$)/$T$ = $\gamma$ + $\beta$$T^2$. 
Note that the present experimental data slightly deviates from the Debye $T^3$ law in the lowest-temperature region, which may be due to magnetic impurities as mentioned above. 
The Debye temperature $\theta_{\rm D}$ can be determined from the following expression: 
\begin{equation} 
\theta_{\rm D} = (12\pi^{4}Rn/5\beta)^{1/3}, 
\end{equation} 
where $R$ is the gas constant and $n$ (= 14) is the number of atoms in the formula unit. 
From the experimental results, we determined $\gamma$ and $\theta_{\rm D}$ as $\sim$ 10.2 mJmol$^{-1}$K$^{-2}$ and $\sim$ 835 K, respectively. 
These obtained values for EuBe$_{13}$ are comparable to those reported for LaBe$_{13}$: $\gamma$ $\sim$ 9 mJmol$^{-1}$K$^{-2}$ and $\theta_{\rm D}$ $\sim$ 750 -- 950 K \cite{Bucher, Hidaka-La, Hidaka-XRD, Kappler}.

\begin{figure}[htb]
\begin{center}
\includegraphics[width=0.7\linewidth]{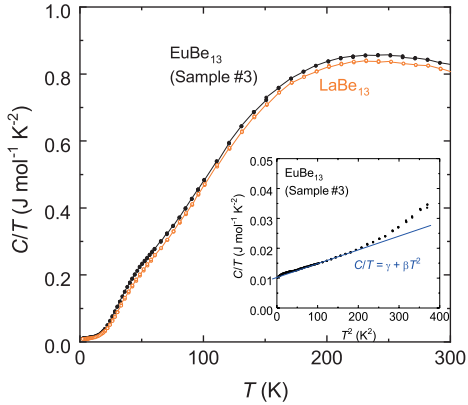}
\caption[]{\protect (Color online) Temperature dependence of $C$/$T$ for EuBe$_{13}$ (closed symbols) and LaBe$_{13}$ (open symbols) \cite{Hidaka-La} below 300 K at zero field. 
The inset shows the low-temperature region of $C$($T$)/$T$ as a function of $T^2$. 
The blue line represents the Debye $T^3$ law. 
}
\end{center}
\end{figure}

\begin{figure}[tb]
\begin{centering}
\includegraphics[width=0.8\linewidth]{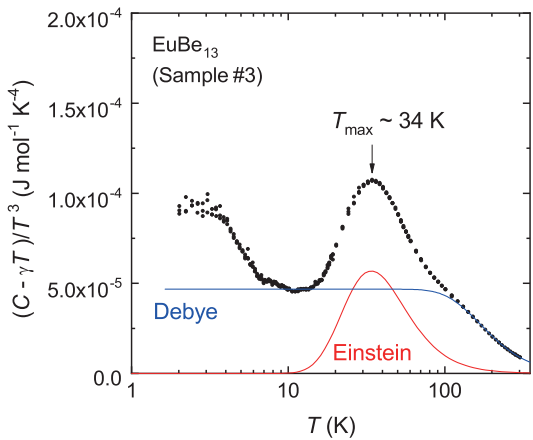}
\caption{(Color online) Temperature dependence of ($C$ -- $\gamma$$T$)/$T^3$ for EuBe$_{13}$. The blue and red curves represent the Debye and Einstein phonon contribution calculated using the obtained $\gamma$, $\theta_{\rm D}$, and $\theta_{\rm E}$.}
\end{centering}
\end{figure}

For nonmagnetic MBe$_{13}$ compounds, a hump structure is observed in $C$($T$)/$T$ at approximately 35 K \cite{Hidaka-La, Felten}, which can be regarded as the contribution of the low-energy Einstein phonon due to the oscillation of the M ion. 
To estimate the contribution of the Einstein phonon, we plotted the temperature dependence of ($C$ -- $\gamma$$T$)/$T^3$ for EuBe$_{13}$ as shown in Fig. 3. 
The blue solid line represents the Debye specific heat calculated using the obtained $\gamma$ and $\theta_{\rm D}$. 
The deviation of ($C$ -- $\gamma$$T$)/$T^3$ from the Debye specific heat below 10 K may be due to magnetic impurities. 
It is noteworthy that the ($C$ -- $\gamma$$T$)/$T^3$ curve shows a broad peak at $T_{\rm max}$ $\sim$ 34 K, which should originate from the contribution of the Einstein phonon. 
$T_{\rm max}$ is linked to $\theta_{\rm E}$ via the relationship $\theta_{\rm E}$ $\sim$ 4.92 $T_{\rm max}$ \cite{Matsuhira}, from which we estimated $\theta_{\rm E}$ for EuBe$_{13}$ to be $\sim$ 167 K.
This estimated $\theta_{\rm E}$ value is fairly close to those reported previously for the other isostructural MBe$_{13}$ compounds, LaBe$_{13}$, SmBe$_{13}$, UBe$_{13}$, and ThBe$_{13}$ \cite{Renker, Hidaka-La, Hidaka-XRD}.

\subsection{Electrical resistivity}

Figures 4(a) and 4(b) display the temperature dependence of the electrical resistivity $\rho$($T$) for EuBe$_{13}$ measured on sample \#1 and LaBe$_{13}$ taken from Ref. 19, respectively. 
The electrical current $\boldmath I$ was applied along the [100] direction in both the measurements. 
The $\rho$($T$) curve for EuBe$_{13}$ exhibits simple metallic behavior and is similar to that for LaBe$_{13}$, indicating that EuBe$_{13}$ is a nonmagnetic metallic compound with Eu$^{3+}$. 
$\rho$($T$) for EuBe$_{13}$ shows $T^3$-like behavior at low temperatures; neither $T^5$ due to the electron--Debye phonon scattering nor $T^2$ due to the electron--electron scattering, as shown in the inset of Fig. 4(a). 
Intriguingly, such $T^3$-like dependence was also observed in $\rho$($T$) for LaBe$_{13}$ [see the inset of Fig. 4(b)]. 
These findings suggest that this unusual temperature dependence is a common feature for nonmagnetic MBe$_{13}$ systems; that is, it originates from the electron--phonon scattering due to the presence of the low-energy phonon mode with $\theta_{\rm E}$ $\sim$ 160 K.

\section{Discussion}

We now consider the following model on the basis of Matthiessen's rule to explain the $\rho$($T$) curves for EuBe$_{13}$ and LaBe$_{13}$: $\rho_{\rm Total}$($T$) = $\rho_0$ + $\rho_{\rm Deb}$($T$) + $\rho_{\rm Ein}$($T$). 
The electron--electron scattering can be ignored since $T^2$ behavior is not observed within the experimental accuracy. 
In this formula, $\rho_0$ is the residual resistivity, $\rho_{\rm Deb}$ is the Debye phonon contribution to $\rho$ described by the Bloch--Gr$\ddot{\rm u}$neisen law written as \cite{Ziman, Omoidensi} 
\begin{equation}
\rho_{\rm Deb}(\it T) = A_{\rm Deb} \Biggl( \frac{T}{\theta_{\rm D}} \Biggr)^{\rm 5} \int_{\rm 0}^{\theta_{\rm D}/T} \frac{x^5}{[\rm exp(\it x) - \rm1][ \rm1 - \rm exp(\it -x)]} dx, 
\end{equation}
\begin{equation}
A_{\rm Deb} = \frac{\it C}{\it m \theta_{\rm D}},
\end{equation}
and $\rho_{\rm Ein}$ is the Einstein phonon contribution to the resistivity written as \cite{Cooper} 
\begin{equation}
\rho_{\rm Ein}(\it T) = \frac {A_{\rm Ein}}{T[\rm exp( \frac{\it T}{\theta_{\rm E}}) - \rm1]([\rm1 - \rm exp( -\frac{\it T}{\theta_{\rm E}})]},
\end{equation}
\begin{equation}
A_{\rm Ein} = \frac{\it KN}{\it m}.
\end{equation}
Here, $C$ is a constant, which is independent of the kind of material, $m$ the mass of an oscillator, $N$ the number of oscillators per unit cell volume, and $K$ a constant, which depends on the electron density of the metal and the electron--local-mode coupling strength.

The $\rho$($T$) curves for EuBe$_{13}$ and LaBe$_{13}$ were analyzed using the above formula of $\rho_{\rm Total}$($T$). 
In a fitting using $\rho_0$, $A_{\rm Deb}$, $\theta_{\rm D}$, $A_{\rm Ein}$, and $\theta_{\rm E}$ as free parameters, we were unable to determine the values of these parameters uniquely, because the obtained values after the fitting depend on the initial parameters. 
Hence, in the present analyses, we fixed $\theta_{\rm D}$ and $\theta_{\rm E}$ to the values determined in other experiments. 
In the case of EuBe$_{13}$, the values of $\theta_{\rm D}$ and $\theta_{\rm E}$ were fixed to those obtained from the present $C$ measurements: ($\theta_{\rm D}$, $\theta_{\rm E}$) = (835 K, 167 K). 
The fixed and obtained fitting parameters are summarized in Table I. 
The calculated $\rho$($T$) curve reproduces the experimental data reasonably well, as shown Fig. 4(a), where the calculated $\rho_{\rm Deb}$ and $\rho_{\rm Ein}$ are also shown. 
On the other hand, for LaBe$_{13}$, we performed the analysis using two sets of $\theta_{\rm D}$ and $\theta_{\rm E}$ as the fixed parameters: ($\theta_{\rm D}$, $\theta_{\rm E}$) = (920 K, 177 K) obtained from the $C$ measurements by the authors' group \cite{Hidaka-La}, named Case 1, and ($\theta_{\rm D}$, $\theta_{\rm E}$) = (820 K, 163 K) obtained from the $C$ measurements by Bucher et al. \cite{Bucher} and the XRD measurements \cite{Hidaka-XRD}, named Case 2. 
In both cases, the $\rho_{\rm Total}$($T$) curves appear to reproduce the experimental data in the main panel of Fig. 4(b). 
However, Case 2 gives a better description of $\rho$($T$) for LaBe$_{13}$ than Case 1 since $\rho_{\rm Total}$($T$) in Case 1 deviates from the experimental data in the $T^3$ plot [see the inset of Fig. 4(b)]. 
The deviation from the experimental data in Case 1 is considered to be due to the rather higher $\theta_{\rm D}$ and $\theta_{\rm E}$ than the typical values reported for the MBe$_{13}$ compounds \cite{Renker, Hidaka-XRD}, although it is unclear why our $C$ measurements for LaBe$_{13}$ give higher values of $\theta_{\rm D}$ and $\theta_{\rm E}$ \cite{Hidaka-La}.

Here, we evaluate the obtained $A_{\rm Deb}$ and $A_{\rm Ein}$ parameters (Table I) from the view point of the oscillators of the Debye and Einstein phonons in MBe$_{13}$ systems. 
Using the $A_{\rm Deb}$ parameters of EuBe$_{13}$ and LaBe$_{13}$ obtained from the present $\rho$ measurements, $A_{\rm Deb}^{\rm EuBe_{13}}$/$A_{\rm Deb}^{\rm LaBe_{13}}$ is estimated to be $\sim$ 1.05. 
When the masses of the Debye oscillators are the same for EuBe$_{13}$ and LaBe$_{13}$, i.e., the Be atom is the Debye oscillator, $A_{\rm Deb}^{\rm EuBe_{13}}$/$A_{\rm Deb}^{\rm LaBe_{13}}$ is calculated to be $\sim$ 0.98 from Eq. (8) using $\theta_{\rm D}^{\rm EuBe_{13}}$ and $\theta_{\rm D}^{\rm LaBe_{13}}$ (Case 2). 
On the other hand, $A_{\rm Deb}^{\rm EuBe_{13}}$/$A_{\rm Deb}^{\rm LaBe_{13}}$ becomes $\sim$ 0.90 on the assumption that the Debye oscillators are the Eu and La ions, which is more distant from 1.05. 
Here, the atomic masses of La and Eu are 138.91 and 151.96, respectively. 
These results support the suggestion that the Be atoms form the crystal lattice described by the Debye model given by the previous powder XRD measurements of MBe$_{13}$ \cite{ Hidaka-XRD}.

For the Einstein phonon, $A_{\rm Ein}^{\rm EuBe_{13}}$/$A_{\rm Ein}^{\rm LaBe_{13}}$ can be rewritten as $m_{\rm Ein}^{\rm LaBe_{13}}$/$m_{\rm Ein}^{\rm EuBe_{13}}$ from Eq. (10) when $K$ and $N$ are the same for the two compounds. 
Here, $m_{\rm Ein}$ is the mass of an Einstein oscillator. 
Since it has been revealed that the Einstein oscillator in the MBe$_{13}$ compounds is the M atom \cite{Renker, Hidaka-XRD}, the masses of La and Eu are adopted as $m_{\rm Ein}^{\rm LaBe_{13}}$ and $m_{\rm Ein}^{\rm EuBe_{13}}$, respectively. 
The estimated $A_{\rm Ein}^{\rm EuBe_{13}}$/$A_{\rm Ein}^{\rm LaBe_{13}}$ ($\sim$ 0.96) shows good agreement with $m_{\rm Ein}^{\rm LaBe_{13}}$/$m_{\rm Ein}^{\rm EuBe_{13}}$ ($\sim$ 0.91). 
This result indicates the validity of the present assumption that the M atom is the Einstein oscillator of the low-energy phonon mode, although we cannot exclude the possibility that the Be atoms are the Einstein oscillators only from this analysis. 
This result also suggests that the conduction electron density and the electron--local-mode coupling of EuBe$_{13}$ are similar to those of LaBe$_{13}$, because we assumed the same value of $K$ for the two compounds in the present analysis. 
To investigate the systematic changes in the parameters of $\rho_{\rm Total}$($T$) and whether the $T^3$-like behavior in $\rho$($T$) are common to the MBe$_{13}$ systems, further studies are required for other MBe$_{13}$ compounds with the nonmagnetic ground state, such as LuBe$_{13}$ and ThBe$_{13}$, are now in progress.

\begin{figure}[tb]
\begin{centering}
\includegraphics[width=0.65\linewidth]{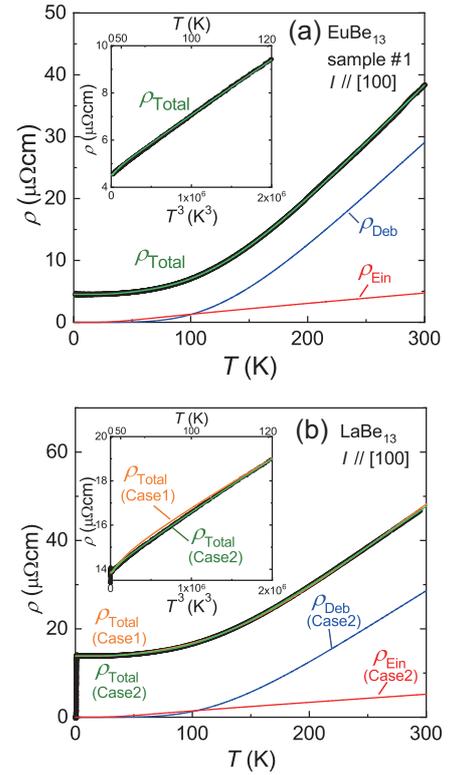}
\caption{(Color online) Temperature dependence of the electrical resistivity of (a) EuBe$_{13}$ measured on sample \#1 and (b) LaBe$_{13}$ taken from Ref. 19. 
The insets in both figures show the $\rho$($T$) data at low temperatures as a function of $T^3$. 
The green and orange solid curves represent the fitting curves of $\rho_{\rm Total}$, as described in the text, while the blue and red solid curves represent the components of $\rho$ attributed to the Debye and Einstein phonon scattering ($\rho_{\rm Deb}$ and $\rho_{\rm Ein}$), respectively. 
}
\end{centering}
\end{figure}

\begin{table*}[htb]
\begin{center}
\caption[]{\protect Fixed parameters of Debye temperature $\theta_{\rm D}$, and Einstein temperature $\theta_{\rm E}$, and fitting parameters of $A_{\rm Deb}$, $A_{\rm Ein}$, and $\rho{_0}$ in the present model calculation for describing $\rho$($T$) of EuBe$_{13}$ and LaBe$_{13}$. 
$\theta_{\rm D}$ and $\theta_{\rm E}$ were taken from the present $C$ measurement and the literature \cite{Bucher, Hidaka-La, Hidaka-XRD}. 
} 
\vspace{0.5cm}
\includegraphics[width=0.9\linewidth]{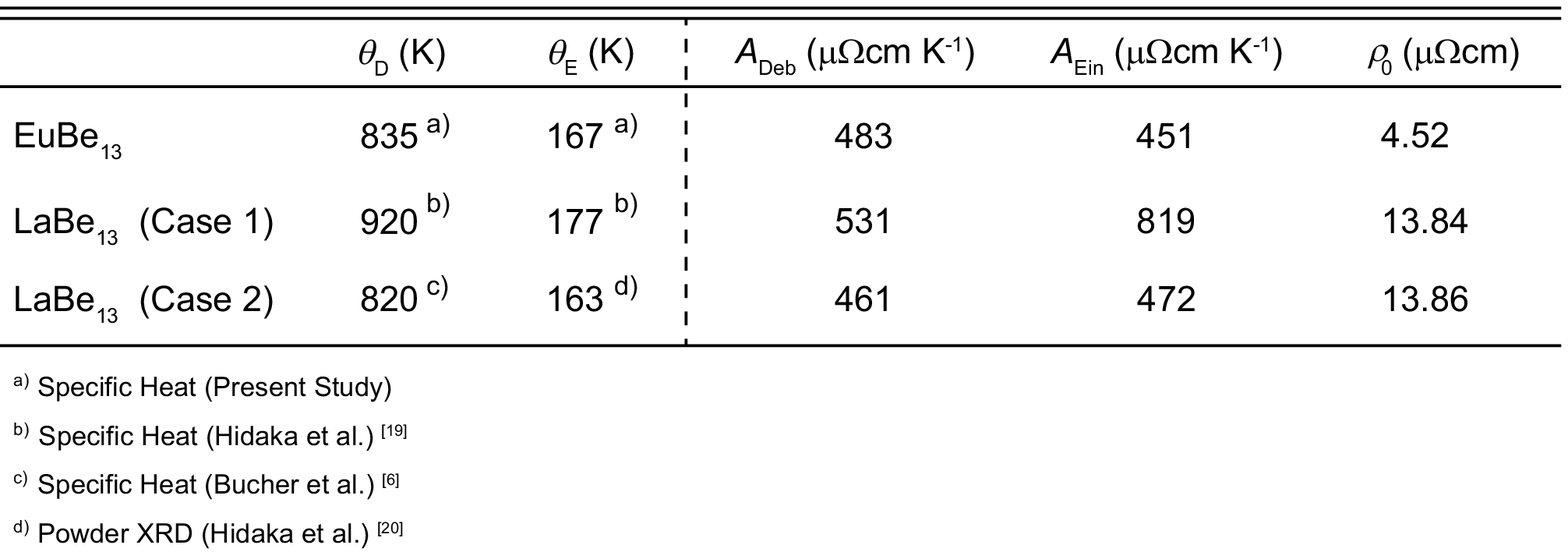}
\end{center}
\end{table*}

Finally we comment on the valence state of the Eu ion in EuBe$_{13}$. 
The valence state of the Eu ion has a strong correlation with the lattice constant because the ionic radius of Eu$^{2+}$ is larger than that of Eu$^{3+}$ \cite{Shannon}. 
In this paper, the free distance of the Eu ion in the material is considered for the comparison among different types of Eu-based compounds. 
Here, the free distance is defined as ${\delta}r$ (= $r_{\rm Eu-FNN}$ -- $r_{\rm FNN }$), where $r_{\rm Eu-FNN}$ is the distance between Eu and the first-nearest-neighbour atom (FNN), and $r_{\rm FNN}$ is adopted to be the covalent atomic radius of the FNN \cite{Covalent}. 
Figure 5 displays the Eu valence states plotted against ${\delta}r$ in various Eu-based cubic intermetallics \cite{Bauer, EuB6, EuFe4P12, EuRu4P12, EuOs4P12, Nakamura-EuX, EuPtSi1, EuPtSi2, EuPtGe}. 
The red and blue solid lines represent the effective ionic radii of Eu$^{2+}$ ($r_{\rm Eu}^{2+}$ = 1.30 A) and Eu$^{3+}$ ($r_{\rm Eu}^{3+}$ = 1.12 A) for the 9-coordination-number site, respectively \cite{Shannon}. 
As seen in Fig. 5, a boundary between Eu$^{2+}$ and Eu$^{3+}$ in these materials appears to be present at ${\delta}r$ $\sim$ 1.8 $\AA$, in other words, there is a tendency for the larger free space to stabilize the Eu$^{2+}$ state. 
Intriguingly, only EuBe$_{13}$ does not follow this tendency. 
Thus, EuBe$_{13}$ is found to be a unique Eu-based intermetallic with the trivalent state in spite of the large free distance.

\begin{figure}[htb]
\begin{center}
\includegraphics[width=0.7\linewidth]{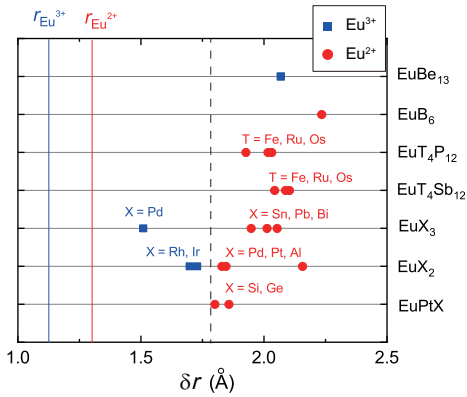}
\caption[]{\protect (Color online) Eu valence states in various Eu-based cubic compounds (circles, Eu$^{2+}$; squares, Eu$^{3+}$). 
${\delta}r$ is a measure of the free distance of the Eu ion. 
The red and blue solid lines represent the effective ionic radii of Eu$^{2+}$ and Eu$^{3+}$, respectively \cite{Shannon}. 
}
\end{center}
\end{figure}

The unique Eu$^{3+}$ state in EuBe$_{13}$ might originate from the characteristic ligands forming the Be-caged structure rather than from the free distance, since the rare-earth MBe$_{13}$ compounds commonly have the trivalent state even in SmBe$_{13}$ and YbBe$_{13}$ \cite{Bucher, Hidaka-Sm, Tsutsui, Heinrich}, except for CeBe$_{13}$ \cite{Wilson, Lenz}. 
One possible explanation is that the ionic radius of the Eu ion is effectively enlarged owing to the low-energy phonon mode with the large local oscillation. 
However, this possibility appears to be unlikely because EuB$_6$ and EuT$_4$Sb$_{12}$ (T = transition metals) also have a low-energy phonon mode, whose $\theta_{\rm E}$ are close to that in EuBe$_{13}$ \cite{Bauer, EuB6-rattling}. 
Another possibility is that the energy loss due to the lattice expansion defeats the energy gain by taking the Eu$^{2+}$ state with the larger ionic radius, even though the present compound has enough free space. 
Since the MBe$_{13}$ compounds have a rigid Be host cage with a high $\theta_{\rm D}$ of $\sim$ 800 K \cite{Hidaka-XRD, Kappler}, the lattice expansion might induce a large energy loss, even for a tiny expansion. 
For comparison, we enumerate $\theta_{\rm D}$ for several La-substituted compounds instead of those for the Eu-based compounds shown in Fig. 5: $\theta_{\rm D}$ = 262 K for LaRu$_4$Sb$_{12}$ \cite{Bauer}, 304 K for LaOs$_4$Sb$_{12}$ \cite{Bauer}, 176 K for LaPd$_3$ \cite{LaPd3}, 205 K for LaSn$_3$ \cite{LaSn3}, 214 K for LaRh$_2$ \cite{LaRh2}, 352 K for LaAl$_2$ \cite{LaAl2}, and 212--885 K for LaB$_6$ \cite{LaB6}. 
Elucidating the origin of the unique trivalent state of EuBe$_{13}$ may provide further insights into not only the $f$ electronic properties in the MBe$_{13}$ systems, including electron--phonon coupling, but also the valence instability in the Eu-based intermetallics.

\section{Summary}
We have succeeded in growing single crystals of EuBe$_{13}$. 
We performed $\chi$($T$), $C$($T$), and $\rho$($T$) measurements on them to investigate the Eu valence state and phonon properties. 
The pure Eu$^{3+}$ state in EuBe$_{13}$ was confirmed by an analysis of $\chi$($T$) on the basis of the Van Vleck theory. 
The contribution of the 4$f$ electrons to $C$ and $\rho$ are negligible below 300 K, and the $C$($T$)/$T$ and $\rho$($T$) curves can be explained well by a combination of the Debye phonon with $\theta_{\rm D}$ $\sim$ 835 K and the Einstein phonon with $\theta_{\rm E}$ $\sim$ 167 K. 
These results indicate that EuBe$_{13}$ is a new member of the MBe$_{13}$ family showing a low-energy phonon mode. 
Furthermore, it is also revealed that the present compound takes an interesting position among the Eu-based cubic compounds with respect to the relationship between the Eu valence and free distance of the Eu ion.

\begin{acknowledgment} 
The authors thank Dr. C. Tabata and Dr. Y. Shimizu for fruitful discussions. 
The present research was supported by JSPS KAKENHI Grants No. JP20224015(S), No. JP25400346(C), No. JP26400342(C), No. JP15H05882, and No. JP15H05885(J-Physics). 
\end{acknowledgment}

\end{document}